\documentclass[preprint,12pt]{elsarticle}

\usepackage{amssymb}

\usepackage{subfigure}

\journal{Journal of Alloys and Compounds}

\begin{document}

\begin{frontmatter}



\title{Density Functional Study of the L1$_0$-$\alpha$IrV Transition
in IrV and RhV}


\author{Michael J. Mehl}

\address{Center for Computational Materials Science, Naval Research
  Laboratory, Code 6390, Washington DC 20375-5000, USA}

\ead{Michael.Mehl@nrl.navy.mil}

\author{Gus L. W. Hart}

\address{Department of Physics and Astronomy, Brigham Young
  University, Provo, UT 84602, USA}

\author{Stefano Curtarolo}

\address{Department of Mechanical Engineering and Materials Science
  and Department of Physics, Duke University, Durham NC 27708, USA}

\begin{abstract}
Both IrV and RhV crystallize in the $\alpha$IrV structure, with a
transition to the higher symmetry L1$_0$ structure at high
temperature, or with the addition of excess Ir or Rh.  Here we
present evidence that this transition is driven by the lowering of
the electronic density of states at the Fermi level of the
$\alpha$IrV structure.  The transition has long been thought to be
second order, with a simple doubling of the L1$_0$ unit cell due to
an unstable phonon at the R point $(0 \frac12 \frac12)$.  We use
first-principles calculations to show that all phonons at the R
point are, in fact, stable, but do find a region of reciprocal space
where the L1$_0$ structure has unstable (imaginary frequency)
phonons.  We use the frozen phonon method to examine two of these
modes, relaxing the structures associated with the unstable phonon
modes to obtain new structures which are lower in energy than L1$_0$
but still above $\alpha$IrV.  We examine the phonon spectra of these
structures as well, looking for instabilities, and find further
instabilities, and more relaxed structures, all of which have
energies above the $\alpha$IrV phase.  In addition, we find that all
of the relaxed structures, stable and unstable, have a density
comparable to the L1$_0$ phase (and less than the $\alpha$IrV
phase), so that any transition from one of these structures to the
ground state will have a volume change as well as an energy
discontinuity.  We conclude that the transition from L1$_0$ to
$\alpha$IrV is probably weakly first order.
\end{abstract}


\end{frontmatter}


\section{Introduction}
\label{sec:intro}

A major goal of computational condensed matter physics is the
determination of structural properties for compounds.  Once the
structure has been determined, other properties, e.g., strength,
ductility, electronic properties (including superconductivity),
magnetism, etc., can be determined.  Several methods exist to attack
this problem.  One possibility is to exhaustively search an
experimental database of known
structures\cite{curtarolo03:datamining}, determining the low energy
structures for each composition of the target materials.  Other
methods use first-principles calculations on a small set of target
structures to determine parameters which can be used to predict
properties of more complex systems.  These include the Cluster
Expansion
Method\cite{sdg84,fontaineCE,azunger:94,axel_CE,uncle},
tight-binding parametrization methods\cite{mehl96:_appli,haas98:tb},
and atomistic potential methods\cite{daw84:eam}.  Methods of the
first type can be combined with methods of the second type for more
through searches\cite{levy10:aflowcem}.

These programs, can, however, only be implemented with a thorough
knowledge of the behavior of compounds.  Exhaustive searches of a
database require an extensive database to search, including all
structures which are known, or thought likely, to form for a target
system.  Parameterized methods may not be able to reach all regions
of phase space, and also need a database of structures for
testing\cite{enum1,enum2}. For both methods, then, it is useful to
look at less-common structures found in nature.

\begin{figure}
  \centering
  \includegraphics[width=9cm]{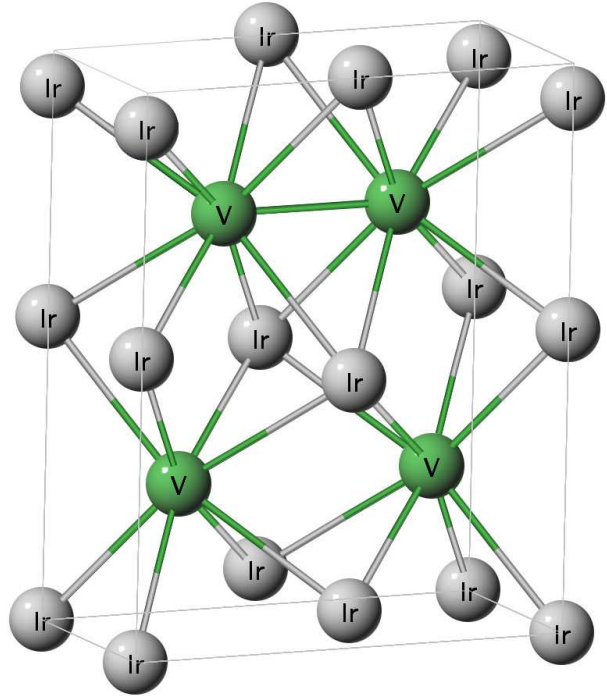}
  \caption{The $\alpha$IrV structure.  The box indicates the
    boundaries of the full orthorhombic unit cell.  In the L1$_0$ or
    (CsCl) structure the Iridium atoms would be in a tetragonal
    (cubic) arrangement around each of the Vanadium atoms.}
  \label{fig:aIrVortho}
\end{figure}

The $\alpha$IrV structure, shown in Figure~\ref{fig:aIrVortho}, is
found in nature only in the prototype
compound\cite{giessen65:aIrV,giessen67:IrV} and its neighbor in the
periodic table, RhV\cite{waterstrat77:RhV,waterstrat73:V3Rh5}. This
orthorhombic structure, space group Cmmm (\#65), can be viewed as a
doubled unit cell distortion of the CsCl
structure\cite{pearson72:crystal} or the tetragonal L1$_0$
structure, space group P4/mmm (\#123)\cite{chen90:IrV}. Indeed, both
the IrV and RhV phase diagrams\cite{baker91:phase} show the L1$_0$
structure as the ground state for Vanadium-poor ($x = 40-48$\%)
[Ir,Rh]$_{1-x}$V$_x$, and Vanadium-rich $\alpha$IrV is known to
transform to L1$_0$ at temperatures above
506$^\circ$~C\cite{chen90:IrV}.

The seemingly straightforward transition pathway from L1$_0$ to
$\alpha$IrV was analyzed by Chen and Franzen\cite{chen90:IrV} in the
context of Landau theory:  First, double the tetragonal L1$_0$ unit
cell along the $y$ and $z$ directions, corresponding to a phonon at
the R $(0\frac12\frac12)$ point in reciprocal space. Second,
construct the primitive base-centered orthogonal unit cell, and
allow the two Ir atoms in the cell to move opposite each other along
the $z$-axis, while the V atoms move along the $y$-axis.  The
resulting structure is symmetrically-equivalent to that of
$\alpha$IrV and, assuming a second-order phase transition, relaxes
into the $\alpha$IrV structure.

This simple picture is not quite accurate.  A true second order
transition, as described by Chen and Franzen, requires a continuous
energy path from L1$_0$ to $\alpha$IrV.  Thus the energy of the
L1$_0$ structure would be lowered by making an infinitesimal
displacement of the type described above, corresponding to a phonon
instability at the R point.  We have performed first-principles
density functional theory calculations for the phonon frequencies at
the R point, both by the frozen phonon method\cite{klein92:pdh} and
by linear
response\cite{baroni87:lrph,baroni01:lrphrev,gonze95:lrph}. We find
that all of the modes here have real frequencies, i.e., they do not
lead to an instability.  

Thus the transition from the L1$_0$ phase to the $\alpha$IrV phase
does not proceed through the simple unit-cell doubling picture
described above.  Instead, as we shall show, the L1$_0$ phase has
vibrational instabilities in another region of the Brillouin zone.
This paper will discuss the structures arising from these
instabilities, and explore the possibility that a second-order
transition might go through one of those phases.

In Section~\ref{sec:elec} we show the results of density functional
theory calculations for the energy and electronic structure of the
$\alpha$IrV and L1$_0$ phases of IrV, RhV, and some neighboring
compounds.  We show that in IrV and RhV the L1$_0$ structure has a
relatively high electronic density of states at the Fermi level, and
so the transition to $\alpha$IrV has Jahn-Teller
character\cite{lee02:bccfcc}.

In Section~\ref{sec:phon} we look at the phonon spectra of the
$\alpha$IrV and L1$_0$ phase.  Not surprisingly, we find that the
$\alpha$IrV phase has no phonon instabilities.  We do, however, find
that the L1$_0$ structure is vibrationally unstable on and near the
line $(x\frac14\frac12)$ in reciprocal space.

In Section~\ref{sec:unstable} we use the frozen-phonon method on the
unstable phonon modes to search for new structures which have lower
energy than L1$_0$.  Such a search is computationally bound, so we
looked at all unit cells with eight atoms or less, and two
structures with 32 atom unit cells.  In this range we find no
instability which relaxes to the $\alpha$IrV phase, however it is
not impossible that searching through larger unit cells would find
such an instability, in which case the transition would indeed be
second order.  We do find several new structures, some of which have
apparently never been seen in nature, and one which was previously
known\cite{waterstrat73:V3Rh5}.

In Section~\ref{sec:disc} we discuss our results, and thoughts on
the order of the L1$_0$-$\alpha$IrV transition.

\section{Energetics and Electronic Structure of the L1$_0$ and
  $\alpha$IrV Phases}
\label{sec:elec}

All computations were made using the
Kohn-Sham\cite{kohn65:inhom_elec} formulation of Density Functional
Theory\cite{hohenberg64:dft} with the
Perdew-Burke-Ernzerhof\cite{perdew96:pbegga} generalized gradient
approximation.  Depending on our needs, we used the Vienna {\em ab
  initio} Simulation Package
(VASP)\cite{kresse93:vasp1,kresse94:vasp2,kresse93:vasp3} with
projector augmented-wave (PAW) potentials\cite{kresse99:ustopaw}, or
the {\em Quantum Espresso} (QE) package\cite{giannozzi09:qe} with
the supplied ultra-soft pseudopotentials for Ir, Rh, and V.

We used rather large plane wave cutoffs of 350~eV in VASP and 540~eV
in QE to ensure convergence.  Similarly, we used $12 \times 12
\times 8$ $\Gamma$-centered k-point meshes in both the L1$_0$ and
$\alpha$IrV structures, leading to 140 and 215 points, respectively,
in the irreducible Brillouin zones.  We summed over electronic
states using a Fermi-Dirac distribution\cite{gillan89:alvac} with a
temperature of 65~meV (0.005 Rydbergs).  These values give
well-converged results.

\begin{figure}
  \centering
  \subfigure{\includegraphics[width=65mm]{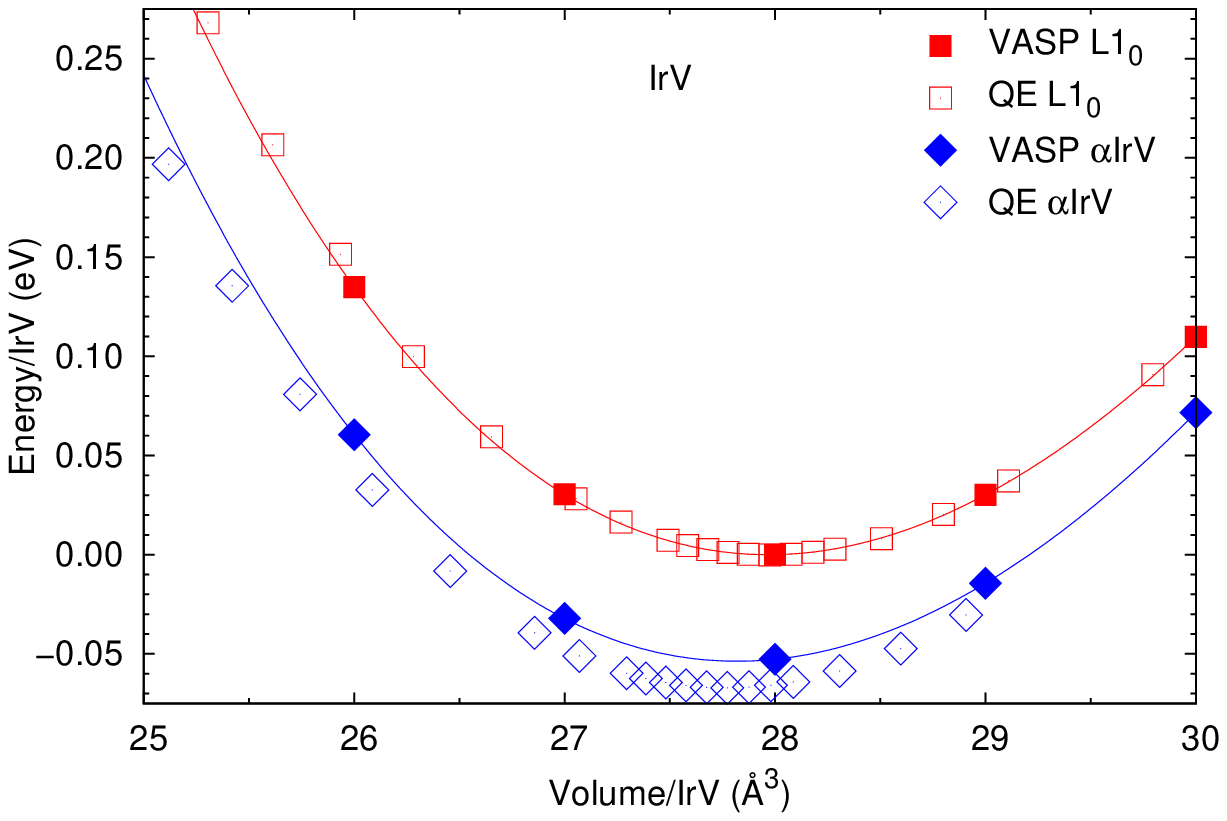}}
  \subfigure{\includegraphics[width=65mm]{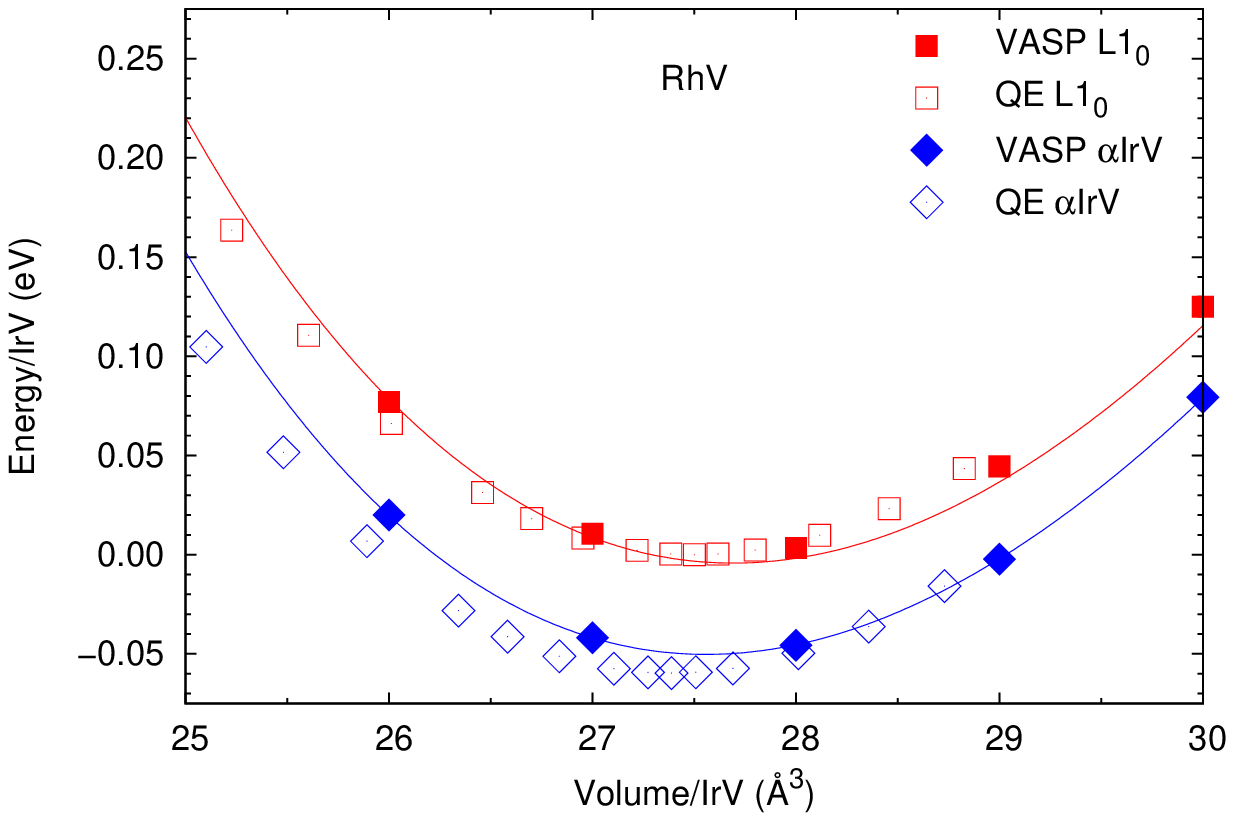}}
  \label{fig:IrVeos}
  \caption{Energy/volume curves for IrV and RhV in the L1$_0$ and
    $\alpha$IrV structures, determined from VASP and QE calculations
    as discussed in the text.  For ease of comparison we have set
    the minimum energy of the L1$_0$ phase to zero.}
\end{figure}

We show our results for IrV and RhV in Table~\ref{tab:structures}
and Fig.~\ref{fig:IrVeos}.  In both cases we see that the VASP PAW
potentials and the QE ultrasoft pseudopotentials are in excellent
agreement with one another, and in good agreement with experiment,
within the usual errors of Density Functional theory in the
generalized gradient approximation.  In both compounds the
$\alpha$IrV state is below the L1$_0$ state by approximately
55~meV/formula unit, in agreement with experiment.

\begin{table}
  \caption{Equilibrium lattice constants (in \AA), atomic positions,
    and equilibrium bulk modulus (K$_0$, in GPa) for the L1$_0$ and
    $\alpha$IrV structures of IrV and RhV determined from
    experiment\cite{giessen65:aIrV,waterstrat77:RhV}, VASP PAW
    calculations, and QE ultrasoft pseudopotential calculations.
    Note that we give the primitive tetragonal lattice parameters
    for the L1$_0$ structure, rather than the more-common
    face-centered tetragonal setting, so that when $c/a = 1$ the
    L1$_0$ structure reduces to the cubic CsCl structure.  The final
    row of the table shows the energy difference between the two
    phases (meV/formula unit).}
  \label{tab:structures}
  \centering
  \begin{tabular}{c|rrr|rrr}
    \hline
 & \multicolumn{3}{c|}{IrV} & \multicolumn{3}{c}{RhV} \\ \hline
    \multicolumn{7}{c}{L1$_0$ Structure} \\ \hline
 & Exp. & VASP & QE & Exp. & VASP & QE \\ \hline
a & 2.749 & 2.755 & 2.762 & 2.754 & 2.739 & 2.745 \\
c & 3.651 & 3.677 & 3.668 & 3.599 & 3.660 & 3.648 \\
K$_0$ & & 272 & 277 & & 230 & \\ \hline
    \multicolumn{7}{c}{$\alpha$IrV Structure} \\ \hline
a & 5.791 & 5.838 & 5.816 & 5.78 & 5.849 & 5.813 \\
b & 6.756 & 6.759 & 6.767 & 6.65 & 6.707 & 6.725 \\
c & 2.796 & 2.814 & 2.823 & 2.78 & 2.792 & 2.802 \\
K$_0$ & & 271 & 276 & 227 & & \\ \hline
Ir/Rh (4j) & 0.22 & 0.216 & 0.216 & & 0.214 & 0.215 \\
V (4g) & 0.28 & 0.296 & 0.297 & & 0.296 & 0.297 \\ \hline
\multicolumn{7}{c}{$\Delta$ E} \\ \hline
 & & 53.6 & 67.0    & & 50.4 & 59.7
  \end{tabular}
\end{table}

\begin{figure}
  \subfigure{\includegraphics[width=65mm]{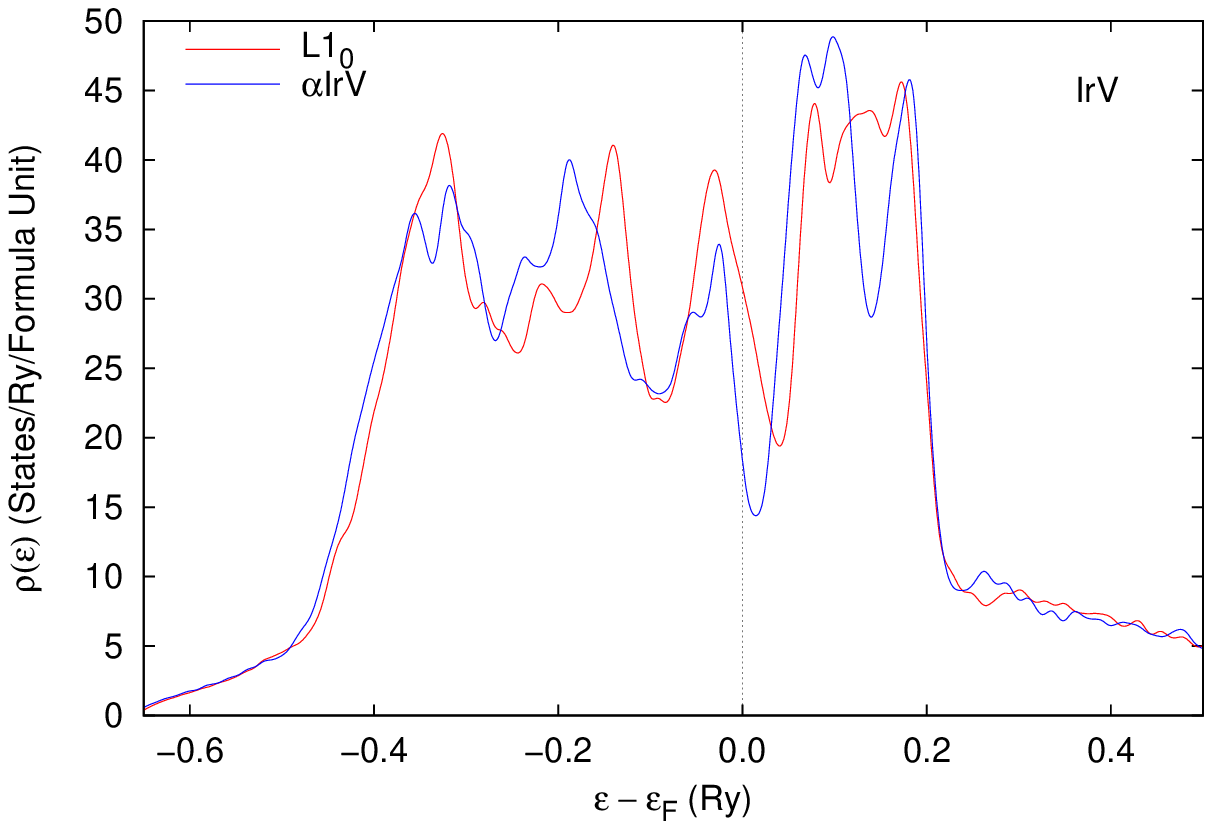}}
  \subfigure{\includegraphics[width=65mm]{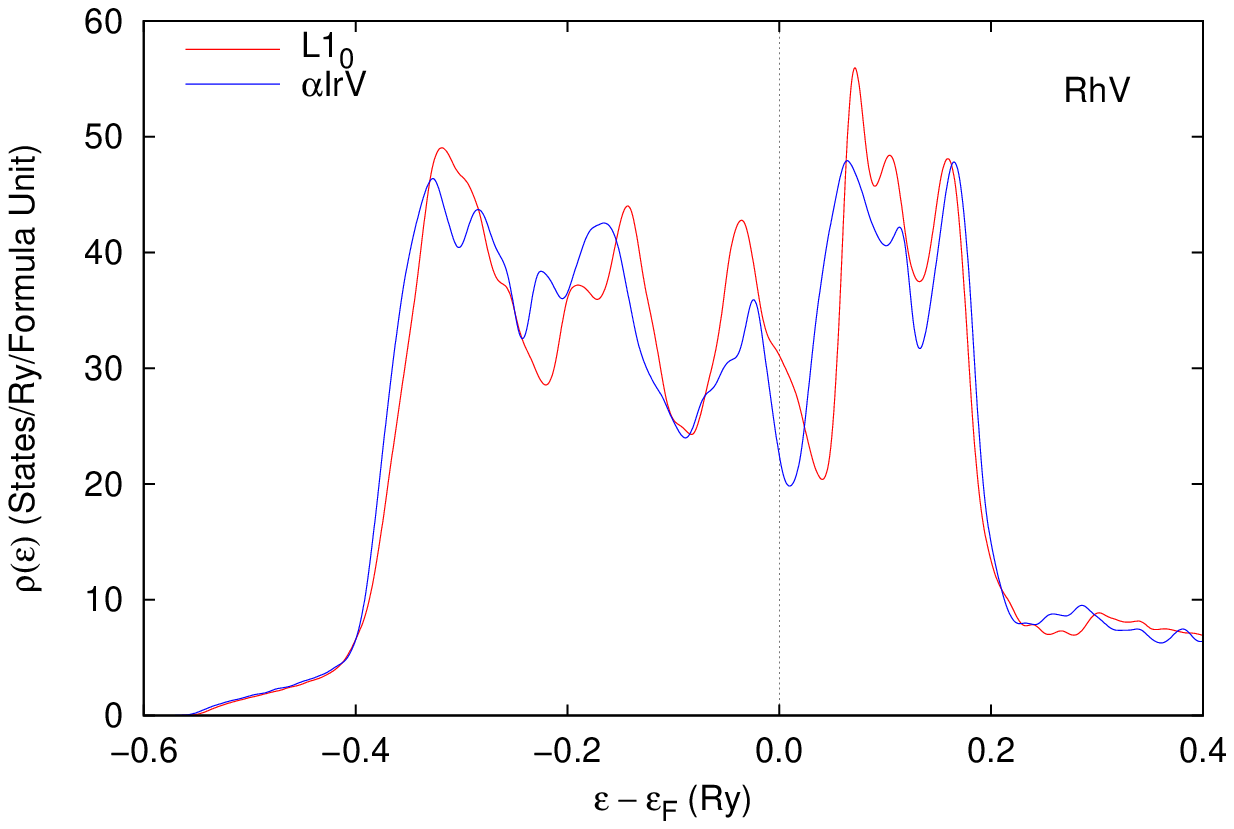}}
  \caption{Electronic density of states for the L1$_0$ (red) and
    $\alpha$IrV (blue) phases of IrV (left) and RhV (right), found
    by smearing the eigenvalues computed by VASP using a Fermi-Dirac
    distribution at T = 65 meV (5
    mRY).\protect\cite{gillan89:alvac}}
  \label{fig:IrVdos}
  \label{fig:RhVdos}
\end{figure}

From experiment\cite{chen90:IrV} we know that L1$_0$ is the
preferred high temperature structure for $\alpha$IrV.  Using the
COM\-SUBS routine from the
ISO\-TROPY\cite{stokes07:isotropy,isodisplace} package, we find that
an orthorhombic distortion of the L1$_0$ structure reduces the
symmetry from space group P4/mmm to Cmmm, the space group of the
$\alpha$IrV structure.  It is plausible to argue\cite{chen90:IrV}
that this lowering of symmetry is the pathway for the L1$_0
\rightarrow \alpha$IrV phase transition.  The symmetry-breaking
character of the transition is evident from Fig.~\ref{fig:IrVdos},
where we plot the electronic density of states of both phases near
the Fermi level.  As we are only interested in the overall behavior
of the density of states, we compute these curves by smearing out
each eigenvalue found by VASP using a Fermi distribution at a
temperature of 5~mRy.  We see that at the Fermi level the density of
states is twice as large in the L1$_0$ phase as it is in the
$\alpha$IrV phase.  This is consistent with a Jahn-Teller-like
symmetry breaking and phase transition\cite{lee02:bccfcc}. As an
aside, we also note that there is a minimum in the L1$_0$ density of
states just above the Fermi level.  This is consistent with the
phase diagram of IrV\cite{baker91:phase}, which shows that
Iridium-rich IrV has the L1$_0$ structure.  Assuming the additional
Ir replaces V in the L1$_0$ structure, and using a rigid-band model,
we can see how adding Ir to the system would raise the Fermi level
and lower the density of states, leading to a more stable L1$_0$
phase.  We find a similar, though less pronounced, state of affairs
in RhV, as is also seen in Fig.~\ref{fig:RhVdos}.

\begin{figure}
  \centering
  \includegraphics[width=90mm]{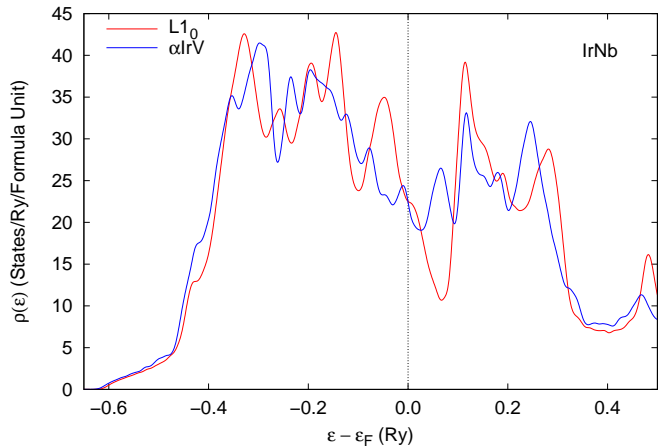}
  \caption{Electronic density of states of the ground-state L1$_0$
    and hypothetical $\alpha$IrV phases of IrNb, found by smearing
    the eigenvalues computed by VASP.}
  \label{fig:IrNbdos}
\end{figure}

We investigated the possibility that the $\alpha$IrV structure might
be found in compounds neighboring IrV and RhV in the periodic table.
The results are shown in Table~\ref{tab:ediff}.  Except for CoPt,
all of the compounds shown here have 14 electrons in the valence
band, just as in IrV and RhV.  In all cases except CoPt we found
that the $\alpha$IrV structure was locally stable, that is, if we
started from the lattice parameters of the $\alpha$IrV, replacing Ir
and V by the indicated atoms, and allowed the system to relax, it
remained in the $\alpha$IrV structure and did not relax to the
higher symmetry L1$_0$ structure.  This suggests that the formation
of the $\alpha$IrV structure is a result of a Fermi level effect
associated with 14 electrons in a tetragonal unit cell.  We did not
check the elastic and vibrational stability of the resulting
structure.  However the L1$_0$ structure was lower in energy for all
compounds except IrV and RhV.  Also, as shown in
Fig.~\ref{fig:IrNbdos}, for these compounds the density of states of
the L1$_0$ structure is very near to that of the $\alpha$IrV
structure, precluding a Jahn-Teller lowering of the energy with
decreased symmetry.  We therefore conclude that it is unlikely to
find the $\alpha$IrV structure anywhere except in IrV and RhV.

\begin{table}
  \caption{Energy difference, in meV per formula unit between the
    L1$_0$ and $\alpha$IrV structures for the given compounds, as
    determined by VASP/PAW/PBE calculations.  A positive number
    indicates that the L1$_0$ structure is lower in energy.  An
    energy difference of zero indicates that the $\alpha$IrV
    structure relaxes into the L1$_0$ structure.  Ground state
    structures are from the {\em ASM Online
    Handbook}\protect\cite{baker91:phase}.
  }
  \centering
  \begin{tabular}{ccccccccc}
Compound & CoPt & CoV & IrNb & IrTa & IrV & RhNb & RhTa & RhV \\ \hline
Ground State & L1$_0$ & D8$_b$ & L1$_0$ & & $\alpha$IrV & L1$_0$ & &
$\alpha$IrV \\ \hline
$\Delta$ E & 0 & 25.3 & 59.0 & 79.9 & -53.6 & 50.1 & 70.6 & -50.4
  \end{tabular}
  \label{tab:ediff}
\end{table}

\begin{figure}
  \centering
  \subfigure{
    \includegraphics[width=70mm]{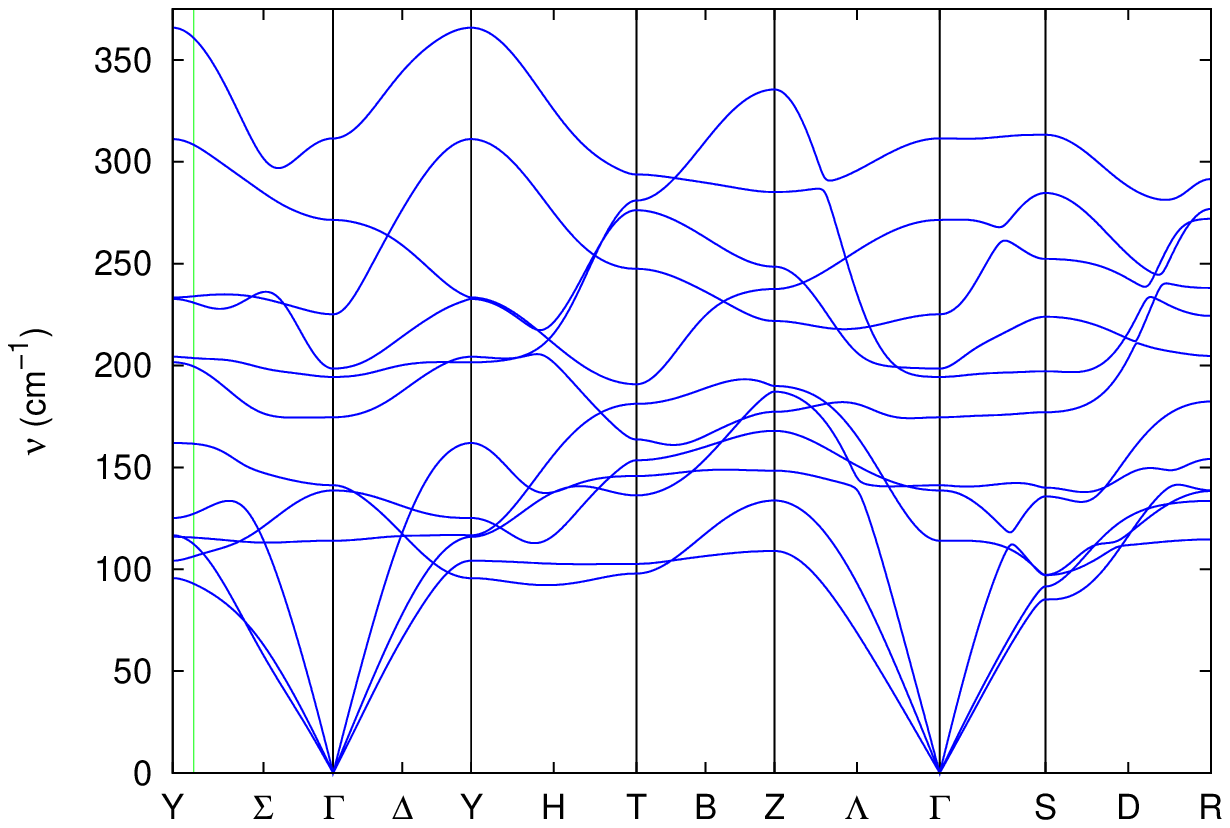}
  }
  \subfigure{
    \includegraphics[height=43mm]{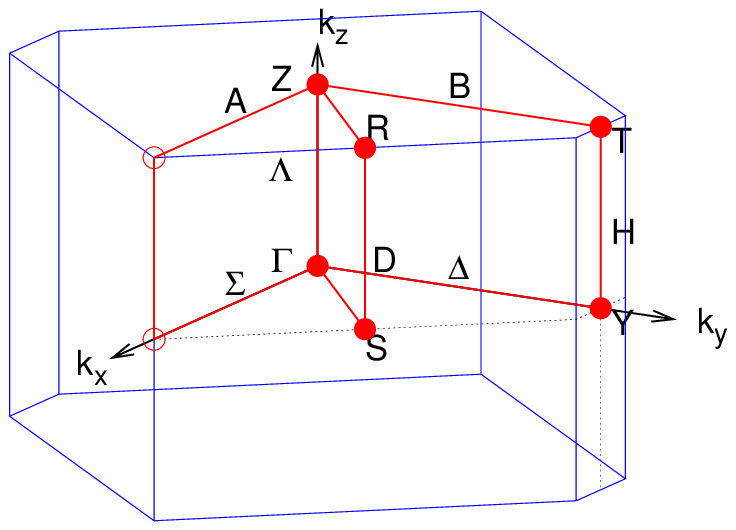}
  }
  \caption{The base-centered orthorhombic $\alpha$IrV structure.
    Left: Phonon frequencies IrV at equilibrium, determined using
    the linear response method with ultrasoft pseudopotentials from
    the {\em Quantum Espresso} package. Right: First Brillouin zone.
    Symmetry lines are described by the notation of Miller and
    Love\protect\cite{miller67:spcgrp}.}
  \label{fig:aIrVph}
\end{figure}

\section{Stability of the L1$_0$ and $\alpha$IrV Phases}
\label{sec:phon}

A system is stable, or metastable, only if it has real phonon
frequencies for all k-points and modes except the acoustic phonons
at the $\Gamma$ point, which are guaranteed to vanish because of
translational symmetry.  Since the elastic constants are related to
the long-wavelength behavior of the acoustic
phonons\cite{kittel96:ssp_cij} this implicitly includes the Born
stability criteria\cite{born66:stability}.  This is fortunate, since
the orthorhombic $\alpha$IrV structure would require us to compute
nine elastic constants.

To check the stability of the $\alpha$IrV phase and the possible
meta-stability of the L1$_0$ phase we computed phonon frequencies
throughout the respective Brillouin zones using linear
response\cite{baroni87:lrph,baroni01:lrphrev,gonze95:lrph}, as
implemented in the {\em Quantum Espresso} package PHon
code\cite{giannozzi09:qe}. Phonon frequencies were computed on a
reciprocal space grid (``q-points'') using an $8\times8\times4$ mesh
for the L1$_0$ structure (45 points in the irreducible Brillouin
zone) and a $6\times6\times 4$ mesh for the $\alpha$IrV structure
(39 points).  Linear response calculations at these points yield a
reciprocal-space dynamical matrix.  These matrices are transformed
into a real-space dynamical matrix which can be used to compute
phonon frequencies over the entire Brillouin zone.

This approach can lead to aliasing at points off of the q-point
mesh.  To check that this does not occur we also performed
frozen-phonon calculations\cite{klein92:pdh}, wherein we construct a
supercell commensurate with the given q-point and measure the change
in energy as a function of atomic displacement within the supercell.
To do this we used the program FROZSL, part of the ISO\-TROPY
package\cite{stokes07:isotropy,isodisplace}, with atomic
displacements of 0.1~Bohr.  Electronic structure calculations were
then performed with the same energy cutoff as the original unit
cell, and when possible the same k-point mesh, albeit folded back
into the smaller Brillouin zone associated with the supercell.

In the following, in interest of saving space, we only show the
results for IrV.  Our calculations for RhV show a similar pattern.

Fig.~\ref{fig:aIrVph} shows the phonon frequencies of the
equilibrium structure of $\alpha$IrV.  High symmetry points and
lines are labeled according to the convention of Miller and
Love\cite{miller67:spcgrp}. All the phonons have real frequencies,
confirming that $\alpha$IrV is at least a metastable structure for
IrV, as suspected.  There is no evidence of any aliasing in these
calculations.

\begin{figure}
  \centering
  \subfigure{
    \includegraphics[width=70mm]{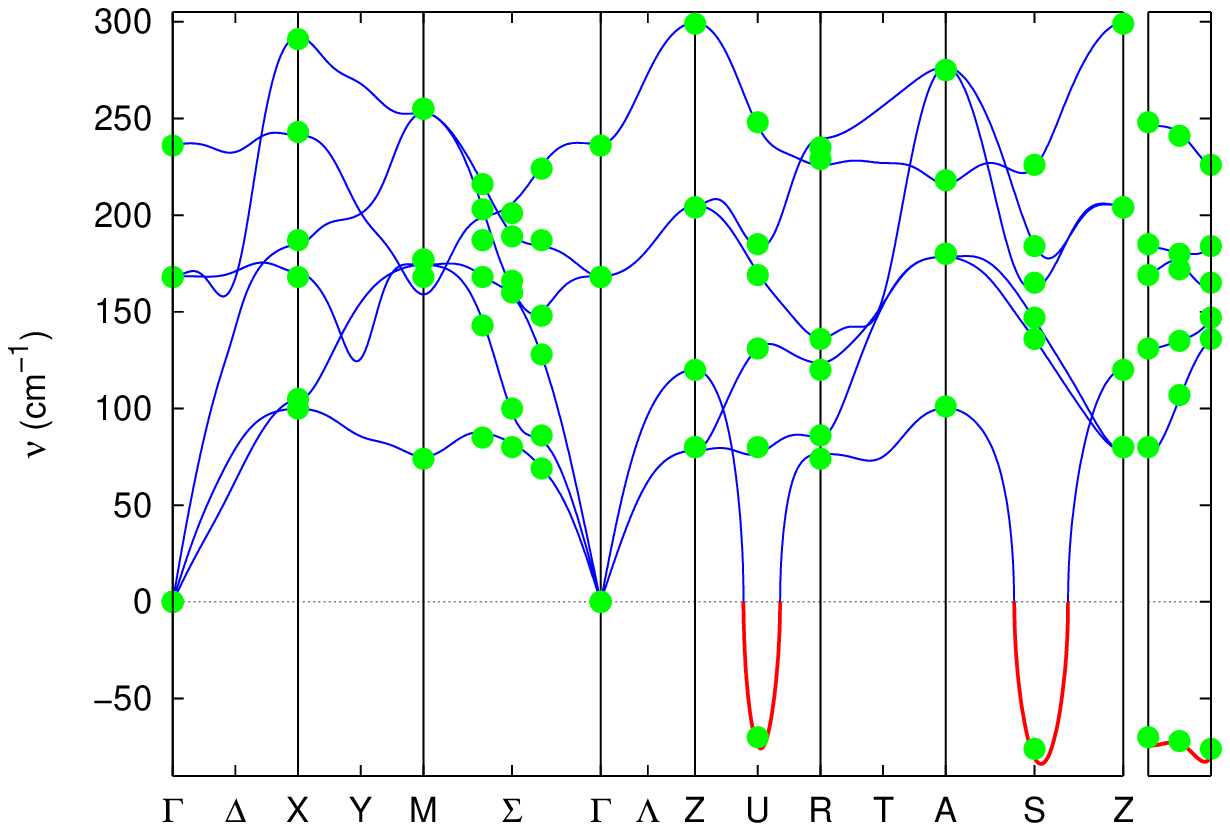}
  }
  \subfigure{
    \includegraphics[width=45mm]{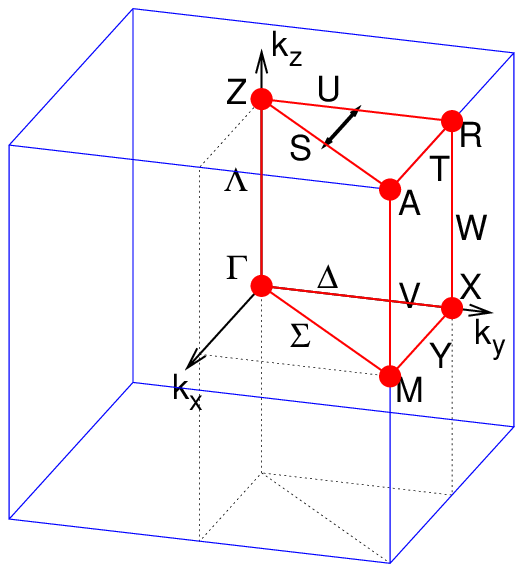}
  }  
  \caption{The tetragonal L1$_0$ structure.  Left: Phonon
    frequencies for IrV at its minimum energy structure, determined
    using the linear response method, using ultrasoft
    pseudopotentials from the {\em Quantum Espresso} package.
    Frequencies below zero (and in red) are unstable modes which
    actually have imaginary frequencies.  The dots represent
    calculations using the frozen phonon method with displacements
    computed by FROZSL.  The second panel shows the frequencies
    along the line $(x \frac14 \frac12)$.  Right: First Brillouin
    zone, with symmetry lines labeled according to the notation of
    Miller and Love\cite{miller67:spcgrp}.  The arrow from U to S
    indicates the region of unstable phonons.}
  \label{fig:l10ph}
\end{figure}

Figure~\ref{fig:l10ph} shows the phonon frequencies of the L1$_0$
phase of IrV, along symmetry lines labeled according to the
convention of Miller and Love\cite{miller67:spcgrp}.  We show
results of frozen phonon calculations, showing good agreement
between the two techniques, even off the linear response q-mesh.  As
expected, there are regions of reciprocal space with imaginary
phonon frequencies.  Surprisingly, this region is not near the R
point $(0 \frac12 \frac12)$ as described by Chen and
Franzen\cite{chen90:IrV}, but along and near a line from the
midpoint of the U line to the midpoint of the S line, as shown in
more detail in the second panel of the figure.  We will discuss the
implications of this instability in the next section.

\section{Searching for Low Energy Structures}
\label{sec:unstable}

The previous section showed that the $\alpha$IrV structure is the
ground state (or, at least, a low-energy metastable state) of IrV,
while the L1$_0$ structure is unstable to phonons along and near the
line $(x \frac14 \frac12)$ in the Brillouin zone, but not at the R
point, where the phonon modes are real, invalidating the
second-order phase transition scenario of Chen and
Franzen\cite{chen90:IrV}.  In this section we will look to see if it
is possible to find another continuous transition path from L1$_0$ to
$\alpha$IrV.

In the frozen phonon method, we displace atoms from their
equilibrium sites in a supercell consistent with the phonon wave
vector and in directions which maintain the symmetry of the phonon
mode.  The frequency of the mode is then directly related to the
square root of the curvature of the energy as a function of atom
displacement.  A mode with an imaginary frequency will then as a
matter of course be related to a supercell calculation which has a
negative curvature, leading to supercells with energy lower than the
original state.  These structures can then be relaxed, leading to
new, or at least different, structures.

As an example of this, consider the unstable L1$_0$ phonon at the
point $(\frac14 \frac14 \frac12)$ on the high symmetry ``S'' line.
The linear response calculations find this phonon frequency to be
81$\imath$~cm$^{-1}$.  To generate a supercell for this calculation
we used the FROZSL code from the ISO\-TROPY
package\cite{stokes07:isotropy,isodisplace}. This package tells us
that this point has phonons in three irreducible representations,
each with two associated modes.  For each representation we run
three total energy calculations for displacements of the atoms in
different directions, with the maximum displacement of 0.1~Bohr.
The energy differences between these structures and the ground state
are used to determine the dynamical matrix at this point, and the
mode frequencies.

For the equilibrium L1$_0$ parameters found in
Table~\ref{tab:structures}, the S$_3$ irreducible representation (in
the notation of Miller and Love\cite{miller67:spcgrp}) has one mode
with an imaginary frequency of 74~cm$^{-1}$.  (The discrepancy
between linear response and frozen phonon calculations is due to the
use of different k-point meshes and the anharmonicity of the mode.
For our purposes we are only interested in showing that the mode is
unstable in both cases, and so will not try to refine the
calculations to improve the agreement between the two.)

Diagonalizing the dynamical matrix allows us to find a supercell
with displaced atoms which has an energy lower than the L1$_0$
phase.  This supercell has space group Cmmm (\#65), and with the
appropriate choice of origin has Ir atoms at the (4e) and (4h)
Wyckoff positions, and V atoms at the (2a), (2b), and (4j)
positions.  This is crystallographically equivalent to the
intermetallic Ga$_3$Pt$_5$ structure~\cite{schubert60:ga3pt5}, and
we will refer to it as such.  Upon relaxation we find a minimum
energy structure which has approximately the same density as, and an
energy 20.5~meV/formula unit below, the relaxed L1$_0$ structure of
Table~\ref{tab:structures}.  Note that this is still well above the
energy of the $\alpha$IrV structure.

In a similar fashion, if we look at the phonons at ${0 \frac14
  \frac12}$, on the ``U'' line, we find an imaginary U$_3$ mode with
a frequency of 70~cm$^{-1}$.  This mode leads to another supercell
with space group Cmmm, but now the Ir atoms are on the (4e) and (4g)
Wyckoff sites, while the V atoms occupy a pair of (4j) Wyckoff
sites.  The relaxed structure again has a density comparable to
L1$_0$, but its energy is only 11.0~meV/formula unit below the
L1$_0$ structure minimum.  In the discussion below will refer to
this structure as Cmmm.

\begin{figure}
  \centering
  \subfigure{
    \includegraphics[width=65mm]{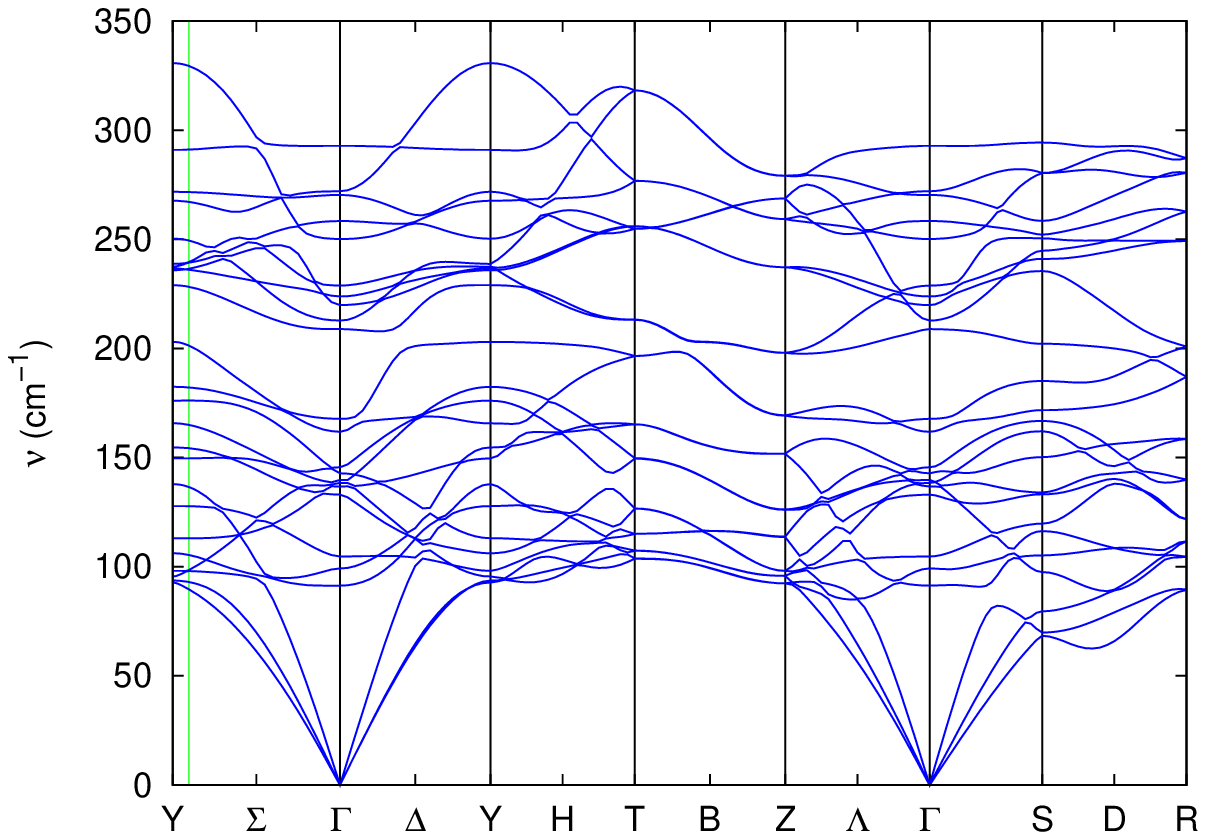}
  }
  \subfigure{
    \includegraphics[width=65mm]{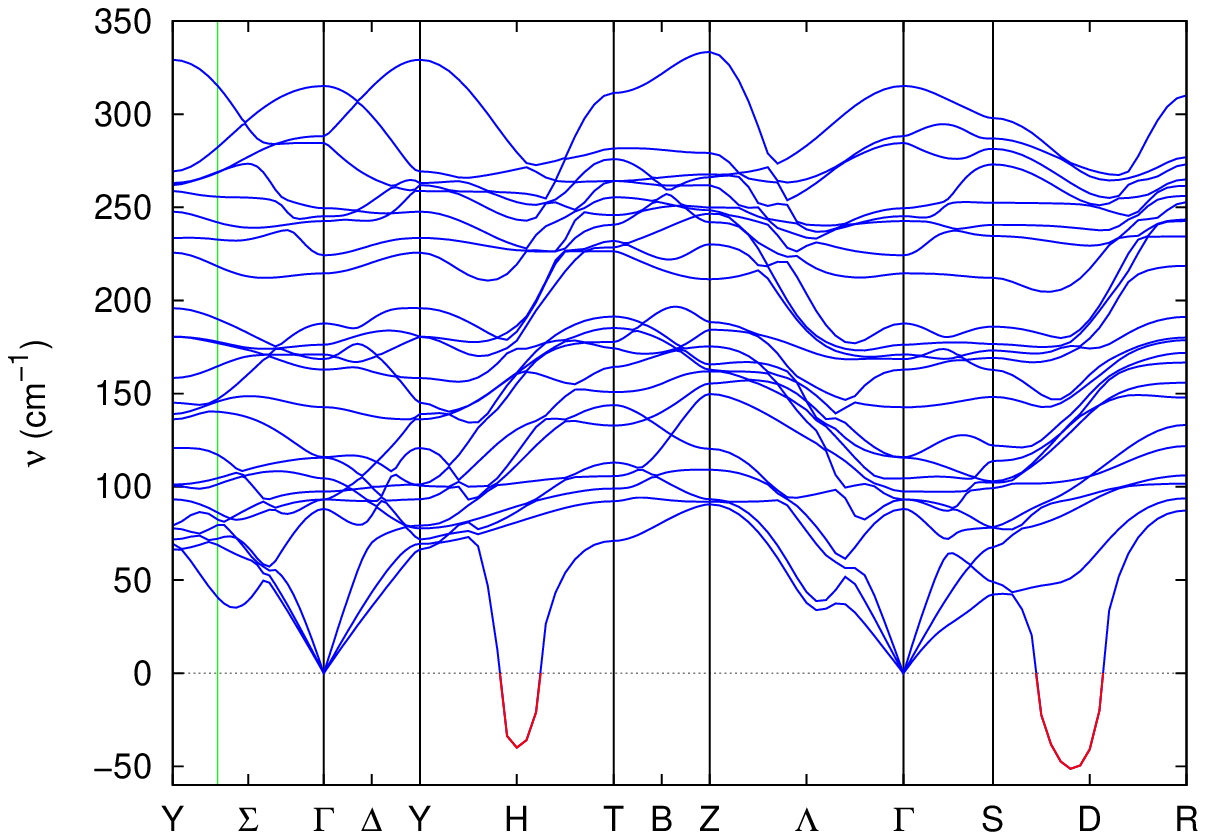}
  }
  \caption{Linear-response phonon spectra of two eight-atom unit
    cells derived from unstable phonon modes of L1$_0$ IrV.  Left:
    The metastable Rh$_5$V$_3$ structure.  Right:  The Amm2
    structure with Ir atoms at two (4d) Wyckoff positions and V
    atoms at four (2b) Wyckoff positions.  The imaginary frequencies
    of the unstable modes are shown as negative frequencies and
    highlighted in red.}
  \label{fig:rh5v3}
\end{figure}

Neither the Ga$_3$Pt$_5$ nor the Cmmm structure will relax to the
ground state $\alpha$IrV structure.  However, both structures have
imaginary frequency long-wavelength optical mode phonons.  If we
venture away from the $\Gamma$ point we will have to deal with
frozen-phonon calculations on unit cells having hundreds of atoms,
so we will only examine the unstable modes at $\Gamma$ in both
structures.

First consider the imaginary frequency $\Gamma_4^-$ mode of the
Ga$_3$Pt$_5$ structure.  It is associated with a supercell of space
group Amm2 (\#38), with Ir atoms at the (4d) and (4e) Wyckoff sites
and V atoms on pairs of (2a) and (2b) sites.  This structure is
crystallographically equivalent to the Rh$_5$V$_3$
structure\cite{waterstrat73:V3Rh5,villars91:rh5v3} and will be
referred to by that name. The relaxed structure is 41.1~meV/formula
unit below the L1$_0$ structure, with similar density.  This
structure is metastable, as seen in the linear-response phonon
spectrum plotted on the left-hand side of Fig.~\ref{fig:rh5v3}.

The second structure, derived from the unstable $\Gamma_3^-$ mode of
Cmmm, again with space group Amm2 (\#38), puts the Ir atoms on two
(4d) sites and the V atoms on four distinct (2b) sites.  The relaxed
structure has an energy 24.1~meV/formula unit below the L1$_0$
structure, again with a similar density.  This structure has not
been found in the intermetallic literature.  It is vibrationally
unstable, as can be seen from the phonon spectrum on the right-hand
side of Fig.~\ref{fig:rh5v3}.

The four structures Ga$_3$Pt$_5$, Rh$_5$V$_3$, Cmmm and Amm2 are the
only structures with eight or fewer atoms in the primitive cell
which can be generated by relaxing the unstable phonon modes of the
L1$_0$ structure.  If we wish to examine larger unit cells, we can
continue along this line of research {\em ad infinitum}, or until we
find a supercell where the atoms relax to the ground state
$\alpha$IrV structure, a supercell with energy {\em below} the
ground state structure, or we run out of computational resources.
In most cases, including this one, we will reach the latter limit
first.  We did compute the relaxed energies of two supercells
associated with unstable phonon modes in the Amm2 structure, and one
with an unstable mode of the Cmmm structure.  These structures had
thirty-two atoms in the supercell, and energies below the parent
structure but above the metastable Rh$_5$V$_3$ structure.

Tables~\ref{tab:energy1}, \ref{tab:energy2} and \ref{tab:energy3}
summarize all of the calculations discussed here, including the
structural derivation, space group, lattice constants, atomic
positions, energy, and stability.

\begin{table*}
  \caption{Structural, energy, and vibrational stability results for
    the structures of IrV discussed in this paper.  All calculations
    use the QE pseudopotentials discussed in the text.  ``Source''
    indicates the origin of the unit cell, either from experiment or
    an unstable phonon in the indicated structure.  ``Atoms'' is the
    number of atoms in the primitive cell.  Lattice constants are
    given in the standard crystallographic convention, e.g. the
    lattice constants of the full
    orthorhombic cell are given for the Cmmm and Cmcm space groups.All
    primitive cells have $\alpha = \gamma = 90^\circ$.  The rows
    ``Ir'' and ``V'' give the Wyckoff positions of the
    atoms. ``Volume'' is the minimum energy volume of the structure
    per formula unit, in (\AA$^3$).  ``Energy'' is the energy of the
    structure below the L1$_0$ structure, in meV/formula unit.}
  \centering
  \begin{tabular}{c|c|c|c}
        \hline
      Structure
    & L1$_0$
    & Cmmm
    & Cmcm\\ \hline
      Source
    & Exp.
    & \parbox{2.2cm}{U$_3 (0\frac14\frac12)$ L1$_0$ phonon}
    & \parbox{2.8cm}{$\Lambda_3 (00\frac14)$ Cmmm phonon} \\ \hline
    \parbox{0.9cm}{Space Group}
    & P4/mmm
    & Cmmm
    & Cmcm \\ \hline
    Atoms
    & 2
    & 8
    & 32\\ \hline
    a (\AA)
    & 2.762
    & 7.260
    & 7.198\\
    b (\AA)
    & 2.762
    & 11.110
    & 11.166\\
    c (\AA)
    & 3.668
    & 2.774
    & 11.136\\
    $\beta$
    & 90
    & 90
    & 90\\ \hline
    Ir
    & (1a) (000)
    & \parbox{2.4cm}{(4e)($\frac14\frac14$0) \\ (4g)(.267 0 0)}
    & \parbox{3.0cm}{(4d)($\frac14\frac14$0) \\ (4e)(.269 0 0)
      \\(4g)(.273 .254 $\frac14$) \\ (4g)(.268 .002 $\frac14$)}
    \\ \hline
    V
    & (1d) ($\frac12\frac12$0)
    & \parbox{2.4cm}{(4j)(0 .364 $\frac12$) \\ (4j)(0 .115
      $\frac12$)}
    & \parbox{3.0cm}{(4f)(0 .358 .134) \\ (4f)(0 .632 .117)
      \\ (4f)(0 .123 .132) \\ (4f)(0 .895 .117)}
    \\ \hline
    Volume
    & 27.97
    & 27.97
    & 27.97 \\ \hline
    Energy
    & 0
    & 10.98
    & 20.61 \\ \hline
    Stability
    & Unstable
    & Unstable
    & Unknown
  \end{tabular}
  \label{tab:energy1}
\end{table*}

\begin{table*}
  \caption{Continuation of Table~\protect\ref{tab:energy1}.}
  \centering
  \begin{tabular}{c|c|c|c}
        \hline
      Structure
    & Ga$_3$Pt$_5$
    & Amm2 
    & Imm2
\\ \hline
      Source
    & \parbox{2.2cm}{S$_3 (\frac14\frac14\frac12)$ L1$_0$ phonon}
    & \parbox{2.2cm}{$\Gamma_3^-$ Cmmm phonon}
    & \parbox{3.5cm}{H$_1 ({\frac14}00)$ \\ Amm2 phonon}
 \\ \hline
    \parbox{0.9cm}{Space Group}
    & Cmmm
    & Amm2
    & Imm2
 \\ \hline
    Atoms
    & 8
    & 8
    & 32
 \\ \hline
    a (\AA)
    & 7.222
    & 2.788
    & 7.176
 \\
    b (\AA)
    & 7.800
    & 7.146
    & 11.144
 \\
    c (\AA)
    & 3.974
    & 11.230
    & 11.190
 \\
    $\beta$
    & 90
    & 90
    & 90
 \\ \hline
    Ir
    & \parbox{2.5cm}{(4e)($\frac14\frac14$0) \\ (4h)(.273 0 $\frac12$)}
    & \parbox{3.0cm}{(4d)(0 .230 .752) \\ (4d)(0 .270 .998)}
    & \parbox{3.5cm}{(4c)(.256 0 .252)\\
      (8e)(.268 2.50 .252)\\
      (4c)(0.223 0 .752)\\
      (4c)(.256 0 .998)\\
      (8e)(.268 .250 .998)\\
      (4c)(.223 0 .498)}
    \\ \hline
    V
    & \parbox{2.5cm}{(2a)(000) \\ (2b)($\frac12\frac12$0) \\ (4j)
      (0 .218 $\frac12$)}
    & \parbox{3.0cm}{(2b)($\frac12$ 0 .352) \\
      (2b)($\frac12$ 0 .625) \\
      (2b)($\frac12$ 0 .125) \\
      (2b)($\frac12$ 0 .899)}
    & \parbox{3.0cm}{
      (4d)(0 .131 .352) \\
      (4d)(0 .381 .355) \\
      (4d)(0 .132 .625) \\
      (4d)(0 .384 .625) \\
      (4d)(0 .130 .125) \\
      (4d)(0 .380 .125) \\
      (4d)(0 .131 .898) \\
      (4d)(0 .381 .895)} \\
 \hline
    Volume
    & 27.98
    & 27.97
    & 27.96
 \\ \hline
    Energy
    & 20.25
    & 24.15
    & 25.12
 \\ \hline
    Stability
    & Unstable
    & Unstable
    & Unknown
  \end{tabular}
  \label{tab:energy2}
\end{table*}

\begin{table*}
  \caption{Continuation of Tables~\protect\ref{tab:energy1}and
    \protect\ref{tab:energy2}.}  \centering
  \begin{tabular}{c|c|c|c}
    \hline
      Structure
    & Cm
    & Rh$_5$V$_3$
    & $\alpha$IrV \\ \hline
      Source
    & \parbox{2.4cm}{$\Delta_1 (0{\frac14}0)$ Amm2 phonon}
    & \parbox{2.4cm}{$\Gamma_4^-$ Ga$_3$Pt$_5$ phonon}
    & Exp. \\ \hline
    \parbox{0.9cm}{Space Group}
    & Cm
    & Amm2
    & Cmmm \\ \hline
    Atoms
    & 32
    & 8
    & 4 \\ \hline
    a (\AA)
    & 13.303
    & 4.070
    & 5.816 \\
    b (\AA)
    & 11.147
    & 7.029
    & 6.767 \\
    c (\AA)
    & 6.652
    & 7.822
    & 2.823 \\
    $\beta$
    & 65.3
    & 90
    & 90 \\ \hline
    Ir
    & \parbox{3.5cm}{
      (2a)(.999 0 .504)\\
      (2a)(.265 0 .974)\\
      (2a)(.485 0 .534)\\
      (2a)(.755 0 .994)\\
      (2a)(.870 0 .256)\\
      (2a)(.140 0 .715)\\
      (2a)(.360 0 .276)\\
      (2a)(.626 0 .745)\\
      (4b)(.992 .249 .521)\\
      (4b)(.260 .252 .983)\\
      (4b)(.365 .748 .267)\\
      (4b)(.133 .251 .729)
      }
    & \parbox{3.0cm}{(4d)(0 .219 .724)\\
      (4e)($\frac12$ .281 .001)}
    & \parbox{2.4cm}{(4j)(0 .216 $\frac12$)}\\ \hline
    V
    & \parbox{3.5cm}{
      (4b)(.178 .130 .344)\\
      (4b)(.175 .380 .362)\\
      (4b)(.312 .134 .625)\\
      (4b)(.312 .385 .625)\\
      (4b)(.064 .125 .117)\\
      (4b)(.060 .375 .133)\\
      (4b)(.447 .130 .906)\\
      (4b)(.450 .380 .888)
      }
    & \parbox{3.0cm}{
      (2a)(0 0 .032)\\
      (2a)(0 0 .457)\\
      (2b)($\frac12$ 0 .218)\\
      (2b)($\frac12$ 0 .793)}
    & \parbox{2.4cm}{(4g)(.297 0 0)}\\ \hline
    Volume
    & 27.97
    & 27.97
    & 27.78 \\ \hline
    Energy
    & 25.67
    & 41.14
    & 67.25 \\ \hline
    Stability
    & Unknown
    & Metastable
    & Ground State
  \end{tabular}
  \label{tab:energy3}
\end{table*}

\section{Discussion}
\label{sec:disc}

We have shown by direct calculation that the transition from the
L1$_0$ to the $\alpha$IrV structure of both IrV and RhV is a result
of a Jahn-Teller driven distortion of the high-symmetry unit cell.
This distortion does not result from a zone-doubling unstable phonon
at the R point of the L1$_0$ Brillouin zone, and so the simple
Landau theory picture does not hold.  This does not completely
eliminate the possibility that the transition is second order, for
we found an entire region of reciprocal space, in and around the
line $(x\frac14\frac12)$, where the L1$_0$ structure has vibrational
instabilities.  We examined several of those instabilities, and
found that they lead to numerous new structures, most of which have
yet to be seen in ordered intermetallic systems and most of which
are vibrationally unstable.  The only phase which we found that is
metastable is the experimentally observed Rh$_5$V$_3$ phase, albeit
in a 50-50 composition.

However, none of the unstable structures shows any sign of relaxing
into the ground state structure.  In fact, looking at
Tables~\ref{tab:energy1}-\ref{tab:energy3}, we see that all of the
structures except $\alpha$IrV have approximately the same volume,
$27.97$~\AA$^3$/formula unit in IrV.  The ground state $\alpha$IrV
structure, one the other hand, has a somewhat smaller volume,
$27.78$~\AA$^3$/formula unit.  While this may seem insignificant,
the volume change from one phase to another is a signal of a
first-order transition, although only weakly first order here.

This paper also shows a mechanism for generating new candidate
intermetallic phases:  look for vibrationally unstable modes in a
high energy structure, construct a supercell which will mimic that
mode within the frozen-phonon calculation, and relax the cell.
Using this method we found five new structures, as well as one,
Rh$_5$V$_3$, which had been seen before but which we had not been
aware of until this research started.  Many more new structures can
undoubtedly be derived from just this system, but further research
in this area is currently restricted by the time needed to search
the Brillouin zone of a given crystal.  For the eight atom
supercells computing the phonon spectrum at 45 q-points with a
reasonable number ($\le 64$) of processors took days to weeks to
complete, depending on the symmetry of the crystal.  Calculating the
phonons for larger unit cells will of course take longer, although
we will have fewer q-points to consider, alleviating some of the
$O[N^3]$ increase due to the larger unit cell.  One of our research
goals will be to construct a ``set-and-forget'' mechanism which will
search the Brillouin zone of an initial structure (here L1$_0$) find
all unstable modes with supercells containing a given number of
atoms, relax those modes, and repeat, until all such structures have
been found.  Even restricting ourselves to all possible binary
intermetallics\cite{curtarolo03:datamining} this will require a
Grand Challenge computational program.

\section{Acknowledgments}
\label{sec:ack}

M. J. Mehl is supported by the United States Office of Naval
Research.  Many of the computations reported here, including all of
the VASP calculations, were performed at the Air Force Research
Laboratory Department of Defense Supercomputing Resource Center,
Wright-Patterson Air Force Base, Dayton OH, under a grant from the
DoD High Performance Computing Modernization Program.  The authors
particularly wish to thank Prof. Harold Stokes for use of his
ISO\-TROPY package.





\bibliographystyle{elsarticle-num}

\bibliography{mehldatabase}

\end{document}